\documentclass[%
 rsi,
 amsmath,amssymb,
 reprint,%
]{revtex4-1}

\usepackage{graphicx}
\usepackage{dcolumn}
\usepackage{bm}
\usepackage{epstopdf}
\usepackage[utf8]{inputenc}
\usepackage[T1]{fontenc}
\usepackage{mathptmx}
\usepackage{etoolbox}

\usepackage{hyperref}

\makeatletter
\def\@email#1#2{%
 \endgroup
 \patchcmd{\titleblock@produce}
  {\frontmatter@RRAPformat}
  {\frontmatter@RRAPformat{\produce@RRAP{*#1\href{mailto:#2}{#2}}}\frontmatter@RRAPformat}
  {}{}
}%
\makeatother
\begin{document}

\preprint{AIP/123-QED}

\title[Review of Scientific Instruments]{Cryogenic Magneto-Terahertz Scanning Near-field Optical Microscope (cm-SNOM)}
\author{R. H. J. Kim}
\affiliation{Ames National Laboratory, US Department of Energy, Ames, Iowa 50011, USA}%
\affiliation{Department of Physics and Astronomy, Iowa State University, Ames, Iowa 50011, USA}%
\author{J.-M. Park}
\affiliation{Ames National Laboratory, US Department of Energy, Ames, Iowa 50011, USA}%
\affiliation{Department of Physics and Astronomy, Iowa State University, Ames, Iowa 50011, USA}%
\author{S. J. Haeuser}
\affiliation{Ames National Laboratory, US Department of Energy, Ames, Iowa 50011, USA}%
\affiliation{Department of Physics and Astronomy, Iowa State University, Ames, Iowa 50011, USA}%
\author{L. Luo}
\affiliation{Ames National Laboratory, US Department of Energy, Ames, Iowa 50011, USA}%
\affiliation{Department of Physics and Astronomy, Iowa State University, Ames, Iowa 50011, USA}%
\author{J. Wang}
\email{jgwang@ameslab.gov; jgwang@iastate.edu}
\affiliation{Ames National Laboratory, US Department of Energy, Ames, Iowa 50011, USA}%
\affiliation{Department of Physics and Astronomy, Iowa State University, Ames, Iowa 50011, USA}%
\date{\today}

\begin{abstract}
We have developed a versatile near-field microscopy platform that can operate at high magnetic fields and below liquid-helium temperatures. We use this platform to demonstrate an extreme terahertz (THz) nanoscope operation and to obtain the first cryogenic magneto-THz time-domain nano-spectroscopy/imaging at temperatures as low as 1.8 K and magnetic fields of up to 5~T simultaneously.   
Our cryogenic magneto-THz scanning near-field optical microscopy, or cm-SNOM, instrument comprises three main equipment: i) a 5 T split pair magnetic cryostat with a custom made insert for mounting SNOM inside; ii) an atomic force microscope (AFM) unit that accepts ultrafast THz excitation and iii) a MHz repetition rate, femtosecond laser amplifier for high-field THz pulse generation and sensitive detection. We apply the cm-SNOM to obtain proof of principle measurements of superconducting and topological materials. The new capabilities demonstrated break grounds for studying quantum materials that requires extreme environment of cryogenic operation and applied magnetic fields simultaneously in nanometer space, femtosecond time, and terahertz energy scales. 

\end{abstract}

\maketitle

The history of modern scientific research stands upon cycles of great discoveries enabled by the development of revolutionary new machines, that allow the study of new states of energy and matter. Each dramatic improvement in operation environment and resolution provides new insights and control of nanostructures in advanced materials. One prominent example of this is the development of scanning near-field optical microscopy (SNOM).   
Recent advances in SNOM techniques have had an exceptional impact on offering nanoscale-resolved images with light well below the diffration limit~\cite{knoll,hill,qazi,chen,fei,hu,nish,hesp}. Implementing near-field techniques give access to local material properties with nanometer spatial precision. For scattering type SNOM; light shines on a sharp metal tip, concentrates electric fields at the end of the tip, and induces polarization charges or free charge carriers on the surface beneath it. The induced carriers follow the oscillations of the illuminating light field, and in turn produces scattered light waves that carry nanoscale information. Thus, the resolution of SNOM is determined by the tip radius, typically orders of magnitude smaller than the wavelength of light, which enable novel ways to control and probe near-field electrodynamic responses in various systems. In particular, there has been considerable progress made toward conducting cryogenic SNOM measurements. Notable examples include the development of home-built systems: to observe the insulator to metal transition of V$_{2}$O$_{3}$ at 150 K~\cite{yang}, image the propagation of plasmon polaritons in graphene at 60 K~\cite{gxni}, and the demonstration of Akiyama-probe-based SNOM at 15 K~\cite{dapo}. There are also works using commercially available SNOM systems that performed successful low-temperature imaging of plasmonic waveguide modes in graphene nanoribbons at 25 K~\cite{zhao} and SNOM demonstration on standard silicon devices down to 9 K at terahertz (THz) frequencies~\cite{hart}.
However, to the best of our knowledge, cryogenic SNOM platforms that allow cooling below liquid-helium temperatures have not been achieved, let alone the application of strong magnetic fields to the samples. Exceeding the current state of art will enable the study of superconducting and topological materials under extreme temperature and magnetic field environments, i.e., a regime that has yet to be accessed. 

\begin{figure}[b]
\includegraphics[width=\linewidth]{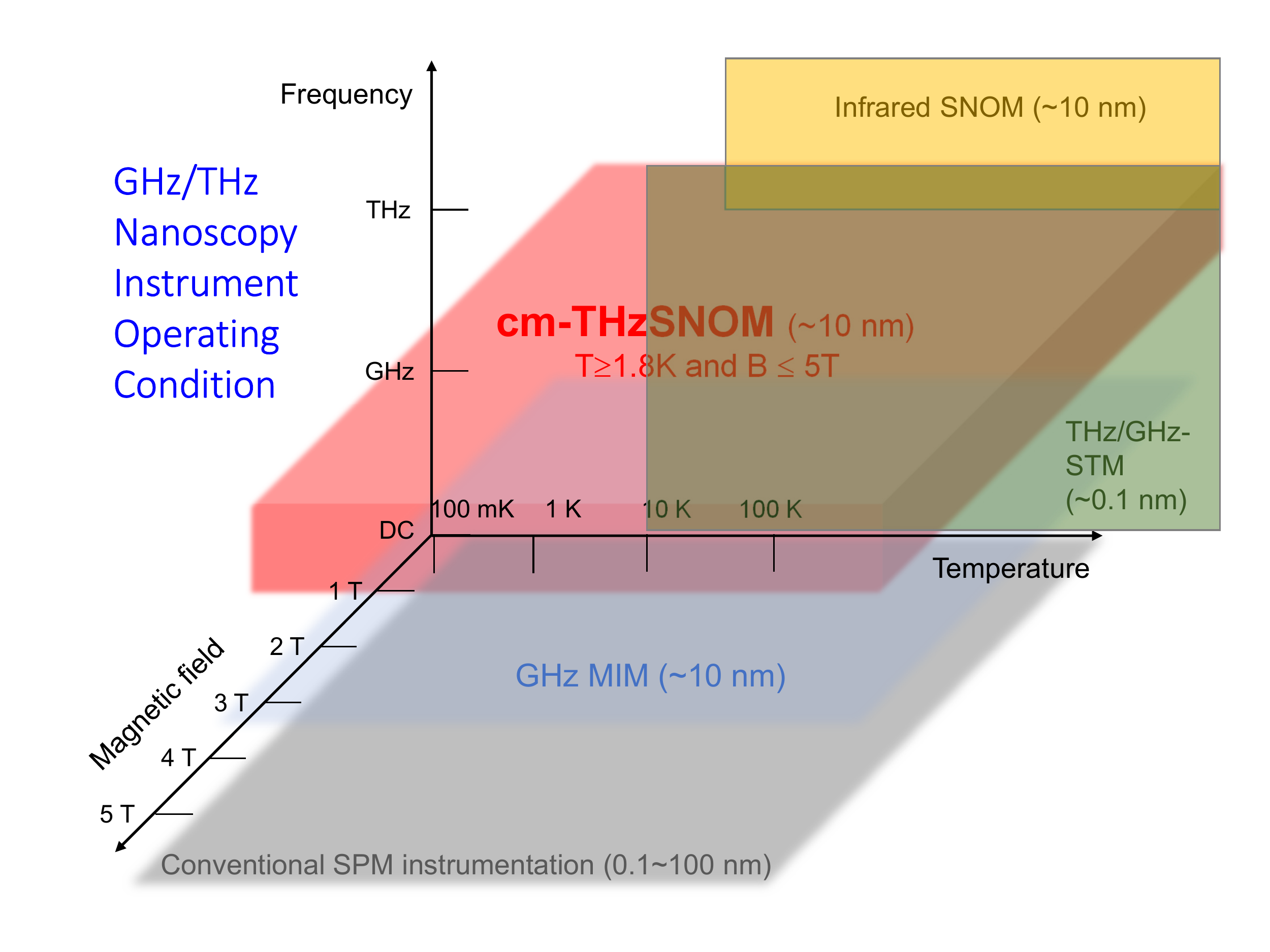}
\caption{\label{fig:1} Comparison of different scanning probe microscopy methods by the energy (frequency) scales that are studied with the temperature and magnetic field conditions that are currently available. The diagram shows regions that each technique can probe within the magnetic field ($x$) --- temperature ($y$) --- frequency ($z$) space. Axes are not drawn to scale. Spatial resolutions are also specified in parentheses for each approach. SNOM: scanning near-field optical microscopy. STM: scanning tunneling microscopy. MIM: microwave impedance microscopy. SPM: scanning probe microscopy.}
\end{figure}

SNOM techniques have recently been extended to light frequencies down to the THz spectral region. By taking advantage of single-/few-cycle THz pulses, it has offered insights to real-space and time-domain characterizations of local electro-optical variations and the emergence of quantum phases at the THz range  \cite{htchen,ribb,zhang,stin,moon,kim,pizz,plank, 1, 2, 3}.
THz-SNOM provides a noninvasive and contactless way to probe the low frequency conductivity of materials at simultaneous nanoscale and sub-picosecond temporal resolution. In particular, the low-frequency THz region is ideal for assessing the fundamental low-energy resonances and dissipationless conductivity peaks toward zero frequency associated with diverse collective modes or exotic topological states in condensed-matter systems~\cite{yang1,yang2,vasw,luo}.

Investigating many fascinating correlation phenomena in solids normally requires magnetic and liquid helium cryogenic operations. In particular, magnetic-field tuning at low-temperature environments is essential to preserve the correlated quantum phases and ground states accessed by GHz to THz frequency probes. 
Although SNOM techniques have achieved tremendous progress to operate along spatial, temporal, or energy (spectroscopic) dimensions, currently available SNOM systems can only be operated at zero-Tesla magnetic field, and above liquid-helium temperatures. This has greatly limited its potential to quantum science and quantum materials.  
For comparison, Fig.~\ref{fig:1} shows dominant nanoscopy tools in terms of operating frequency, cryogenic temperature, and magnetic field. Conventional scanning probe microscopy (SPM), such as AFM and scanning tunneling microscopy (STM), operates only at DC or zero frequency (gray $x$--$y$ plane). Microwave impendence microscopy (MIM) can operate under magneto-cryogenic environments but only at 5$\sim$10 GHz; that is far below correlation gaps of quantum materials (light blue plane). On the other hand, infrared SNOM and THz/GHz-STM~\cite{cock,wang} are operated at zero magnetic field (yellow and dark green boxes in the $x$--$z$ plane). 
As we show below, our cm-SNOM system fills in the unexplored territory with simultaneous THz frequency and nanometer space operation up to 5 T magnetic field and down to liquid-helium temperatures (red box). Our system uniquely fills the empty phase space and, thereby, complements to the existing powerful nanoscopy techniques. 

\begin{figure*}[t]
\includegraphics[width=\linewidth]{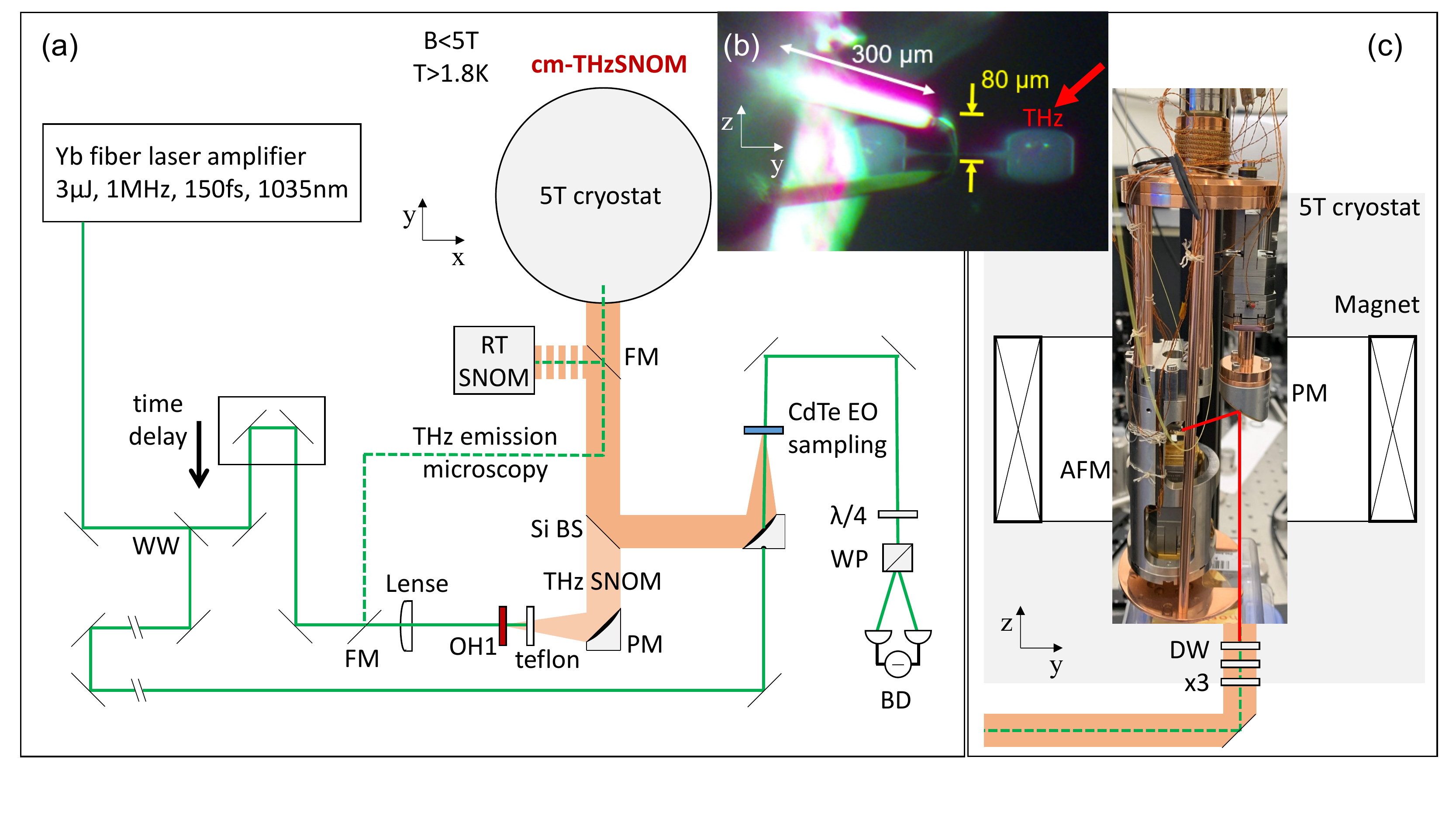}
\caption{\label{fig:2} (a) Experimental setup for THz cryogenic magneto-SNOM. (b) CCD image of the AFM tip approached to a nanostructured sample. The length of the tip and cantilever are around 80 and 300 $\mu$m, respectively. Their reflections from the sample surface are also visible. The THz beam shines the tip and sample area from the top right corner. (c) The AFM unit and the off-axis parabolic mirror on a three-stack piezo stage attached to the cryostat sample probe. When inserted into the sample space of the cryostat, this core piece centers around a split pair magnet that generates a magnetic field along the $z$-direction (perpendicular to the sample surface). WW: wedged window. FM: flip mirror. PM: parabolic mirror. BS: beam splitter. WP: Wollaston prism. BD: balanced detector. DW: diamond window.}
\end{figure*}

Here we establish a cryogenic magneto-THz scanning near-field optical microscopy (cm-SNOM) instrument for dramatic gains in operation environment and resolution by developing a versatile platform that demonstrates the first THz-SNOM operation down to 1.8 K with the ability to apply magnetic fields up to 5 T.  
We discuss the design and development of the cm-SNOM instrument, its performance at room temperature, as well as the successful THz-SNOM demonstration at extreme cryogenic temperatures and magnetic fields in superconducting and topological materials.
Our cm-SNOM technique opens materials observation at nanometer length scales with the tuning capabilities in the magnetic field ($x$) --- temperature ($y$) --- frequency ($z$) measurement space as shown in Fig.~\ref{fig:1}. This machine especially complements magnetic-field-compatible SPM tools that evaluate static or DC electromagnetic properties, or MIM~\cite{barb} that is also limited to isolated narrow bands of microwave frequencies. 

Our THz cm-SNOM is a custom-built instrument based on a MHz repetition rate femtosecond laser amplifier, a 5 T split pair magnet cryostat, an insert for mounting/controlling SNOM inside and outside, and an AFM unit. As shown in Fig.~\ref{fig:2}(a), a femtosecond ytterbium fiber laser with a pulse energy of 3 $\mu$J, repetition rate of 1 MHz, and central laser wavelength of 1035 nm is used to generate ultrafast THz pulses in a nonlinear THz-emitting OH1 organic crystal. The THz signals are detected by electro-optical sampling in a CdTe crystal. This customer-made THz setup is able to achieve a SNR of 100$\sim$120 dB for THz power detection.    
A room-temperature testbed, referred as to RT-SNOM, that mimicks the inside of the cryostat has been used to check beam alignments toward the parabolic mirror adjacent to the AFM. The laser beam can also be guided along a separate path that bypasses the THz generation optics to directly excite the sample to perform THz emission microscopy. The AFM incorporated here is a cantilever-based tapping-mode system (attocube systems AG) designed for applications at low temperature and in high magnetic fields. The metallic AFM tip (Rocky Mountain Nanotechnology, LLC), as shown in Fig.~\ref{fig:2}(b), plays the role of an antenna that receives and transmits far-field THz radiation and amplifies the near-field interaction through a resonant enhancement of the THz field as the length of cantilever and tip roughly matches the THz wavelength. The cryostat is a top-loading, 5 Tesla split pair magnet wet cryostat (ICEoxford Ltd) with a base temperature of 1.7 K. Inclusion of the AFM, extra piezo stacks to control the parabolic mirror, and wire connections slightly raises the lowest SNOM measurement temperature to 1.8$\sim$1.9 K. The helium exchange gas method used inside the sample tube presents a convenient way to cool the entire AFM unit including the sample and replaces a cold finger device that sometimes demands a more laborious process to ensure thermal connections. As illustrated in Fig.~\ref{fig:2}(c), the superconducting magnet surrounds the sample space and supplies a static magnetic field perpendicular to the sample surface. Diamond windows, which offer a wide transparancy window in the electromagnetic spectrum, have been implemented for potential SNOM studies including mid- to near-infrared or visible wavelengths. 

\begin{figure}[]
\includegraphics[width=\linewidth]{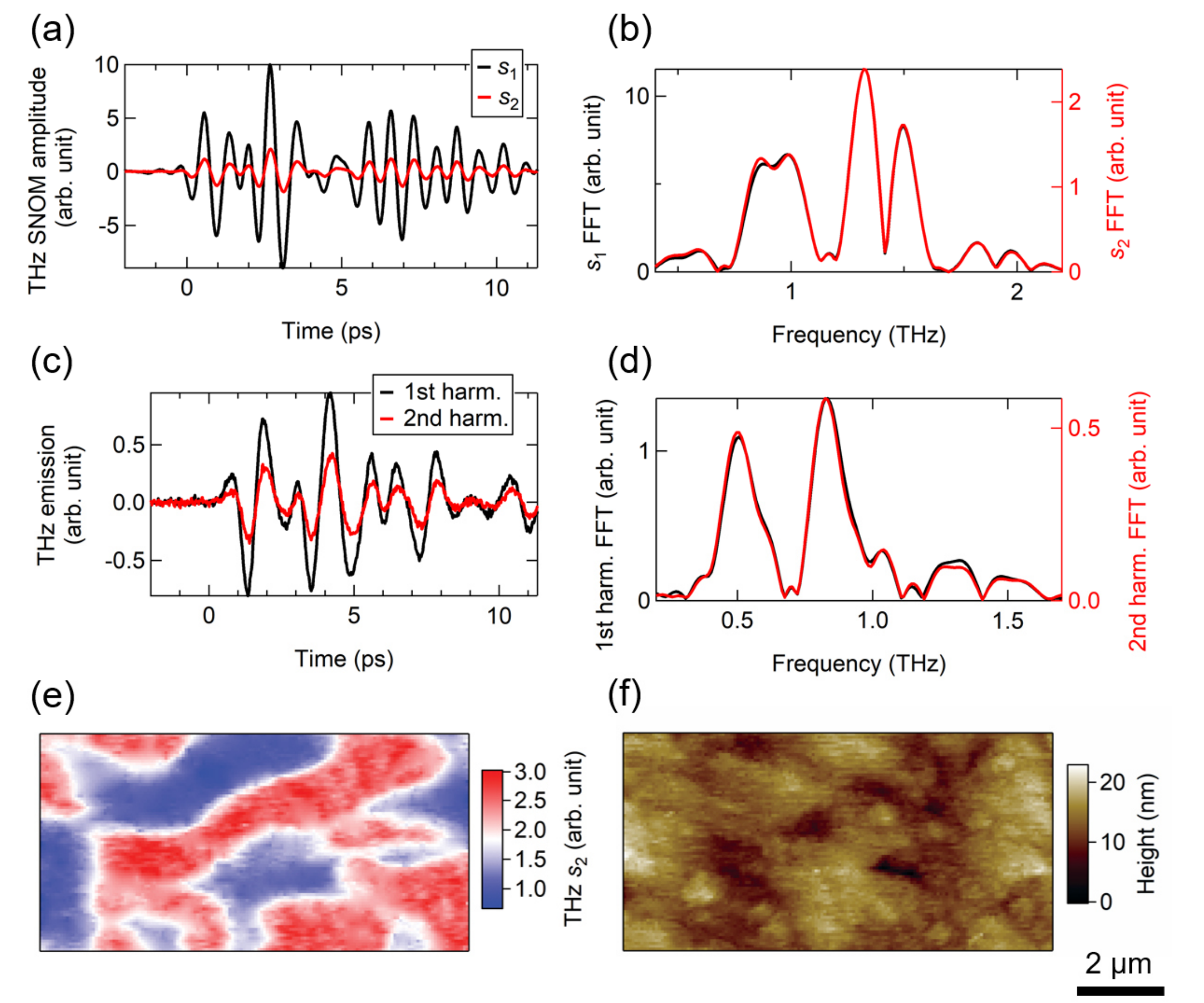}
\caption{\label{fig:3} Room temperature THz near-field measurements. (a) Temporal waveforms and (b) its fast Fourier transformed spectra of the scattered THz near-field amplitude from gold at the first ($s_{1}$) and second ($s_{2}$) harmonic of the tip tapping frequency. (c) Time-domain waveforms and (d) Fourier-transformed spectra of laser-induced THz emission from a InAs single-crystal substrate measured at the first and second harmonic of the tapping oscillation. (e) Spatial map of THz $s_{2}$ and (f) simulaneously acquired AFM topography map of a representative region of a La--Si alloy. The conductivity contrasts in the THz SNOM image reveals the microstructures in the alloy system.}
\end{figure}

\begin{figure}[]
\includegraphics[width=\linewidth]{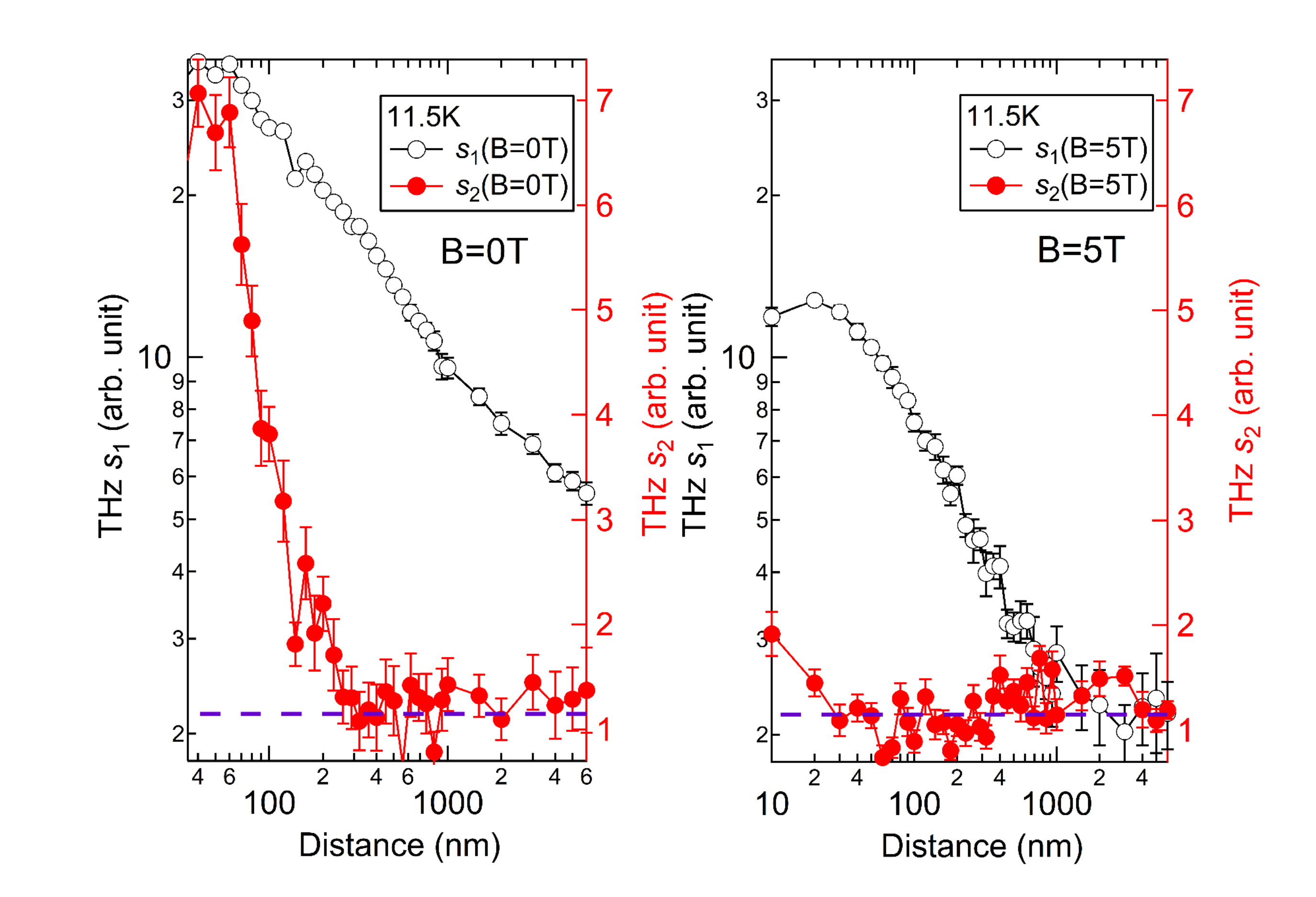}
\caption{\label{fig:5} Low-temperature ($T$ = 11.5 K) approach curves demonstrated on a topological semimetal ZrTe$_{5}$ at 0 T (a) and 5T (b) magnetic fields. The tip-scattered THz signal $s_{1}$ and $s_{2}$ is plotted as a function of the tip-sample distance. For each tip-sample distance, a mean of 20 data points were taken, and the error bars extend plus and minus the average up to the standard deviation.}
\end{figure}

Using the separate RT-SNOM platform outside the magneto-optical cryostat, we have conducted SNOM reference tests on a gold film and THz emission nanoscopy on a InAs substrate. 
The RT-SNOM testbed follows the exact same design used inside the cryostat so that it can serve as both an alignment guide and as a reference to gain insights on the signal-to-noise performance of our THz cm-SNOM when engaged in low-temperature experiments. Once the tip is in contact and positioned at a fixed location on the sample, time-domain THz spectroscopy can be performed by moving the motorized stage that controls the time delay of the optical sampling pulse to the electro-optic crystal and thus traces out the oscillating electric-field waveform of the scattered THz near-field amplitude. Near-field signals $s_{n}$, as shown in Fig.~\ref{fig:3}(a), are then extracted from the scattered THz signal by demodulating the backscattered radiation collected from the tip-sample system at $n$-th harmonics of the tip-tapping frequency of the AFM ($n$ = 1, 2, etc.). A Fourier transform of the time-domain signal gives the spectrum that broadly spans from 0.5 to 2 THz as shown in Fig.~\ref{fig:3}(b). There are noticable cuts in some parts of the spectrum which is also represented by the long ringings continued in the time trace. These are artifacts that may slightly vary with the beam alignment and are a product of the backscattering geometry of our optics. A suitable time domain range and windowing function are applied before the Fourier transform of temporal traces. 
Fig.~\ref{fig:3}(c) and (d) displays THz emission pulses scattered from the tip on an InAs substrate with a laser pulse energy of less than 50 nJ. Interestingly, the emitted signals show a larger ratio of 2$^{\mathrm{nd}}$ to 1$^{\mathrm{st}}$ harmonic signal than in SNOM and presents a spectrum that is skewed toward the lower THz frequencies due to the differing THz generation mechanism in InAs. Intriguingly, the 1$^{\mathrm{st}}$ and 2$^{\mathrm{nd}}$ harmonic signals for both the SNOM and emission microscopy show the very similar waveform and spectrum, which alludes that THz $s_{1}$ are also dominantly near-field and could be employed in circumstances where higher harmonic near-field signals are faint and difficult to attain. These conclusion is corroborated by the approach curves measured in Fig.~\ref{fig:5}. 

Next, to obtain near-field images, the sample stage underneath the tip was raster scanned while the THz sampling delay is fixed to a position that gives the largest near-field amplitude. In Fig.~\ref{fig:3}(e) and (f) we show the THz-SNOM image results from a rare-earth alloy of lanthanum silicide (La--Si). In this alloy, the higher levels of scattered signals (red) represent regions with higher atomic percentage of metallic lanthanum and the lower levels of scattered signal (blue) indicate areas where the relatively less conductive silicon is more abundent. The AFM image of the surface topography that is taken simultaneously with SNOM shows a rather flat surface within a 20 nm roughness. Thus, the THz-SNOM image of the alloy microstructures manifests conductivity contrasts from the near-field Drude response and can be used to extract the local varations in alloy compositions. Accordingly, near-field THz imaging offers an independent pathway to study samples with nanoscale spatial resolution, and the single-cycle THz source further allows the possibility to examine sub-picosecond time dynamics in these measurements.

\begin{figure}[]
\includegraphics[width=\linewidth]{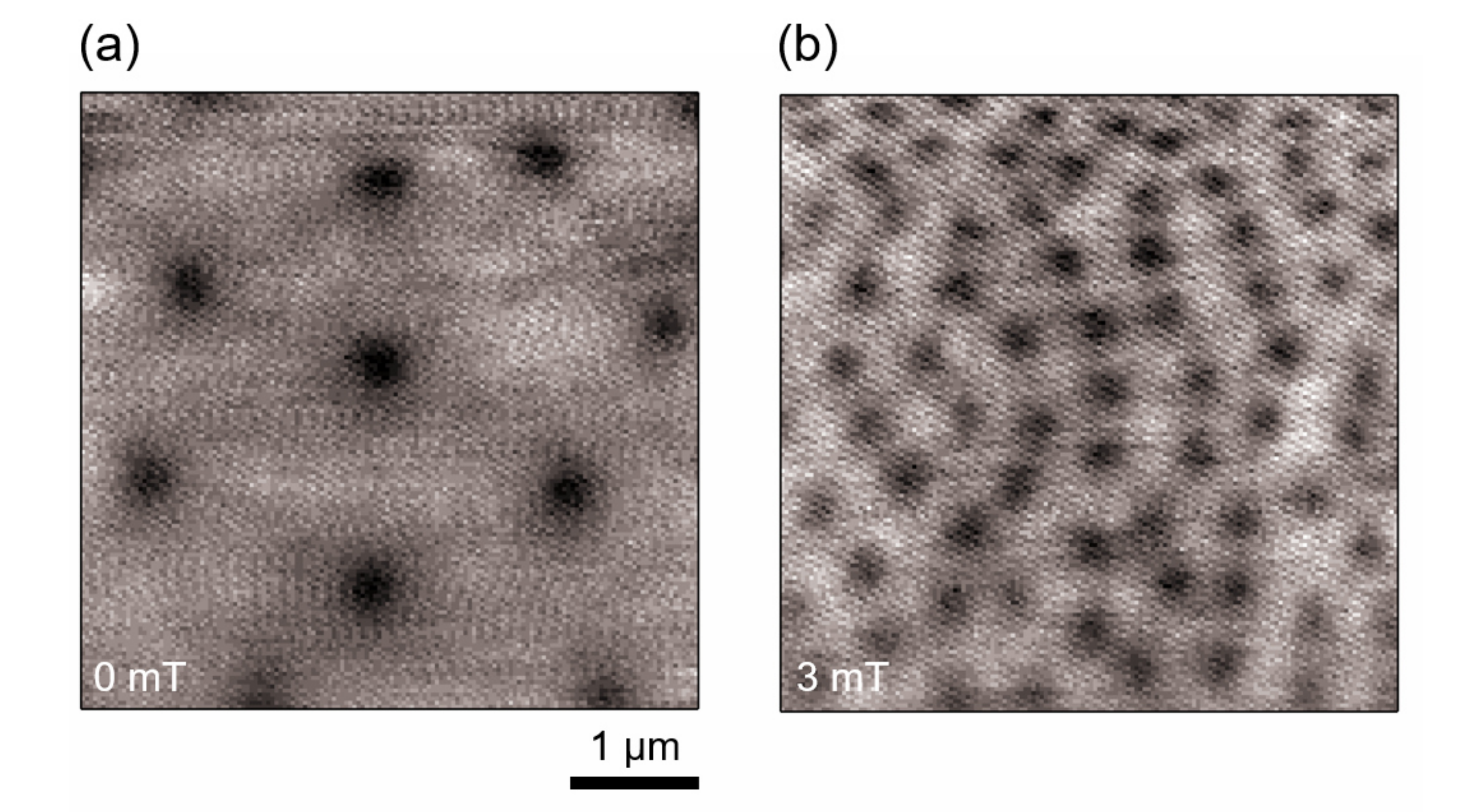}
\caption{\label{fig:4} MFM images of vortices on a 155-nm-thick Nb film measured at 1.8 K with a scan height of 120 nm in an applied magnetic field of (a) 0 mT and (b) 3 mT.}
\end{figure}

Prior to light-coupled SNOM measurements, we carried out magnetic force microscopy (MFM) to image magnetic vortices in superconducting states as a first step to confirm our cryogenic AFM apparatus and characterize lowest temperatures achievable. We use a cobalt-coated magnetized tip on a 155-nm-thick niobium (Nb) thin film deposited on a silicon substrate. The film is known to have a $T_{\mathrm{c}}$ of $\sim$9 K at which it becomes superconducting. In MFM, the changes in the response of the oscillating tip that is lifted above the sample surface, by approximately 120 nm, records the variations in the magnetic field across the sample plane. As depicted in the images of Fig.~\ref{fig:4}(a) and (b), an array of dense vortices occurs by magnetic fields penetrating into the superconductor at an applied field of 3 mT and field-cooled down to 1.8 K. Multiplying the approximate 50 vortices shown in the MFM image by the magnetic flux quantum (= $h/2e$) exactly matches the product of the scanned area times the magnetic-field strength of 3 mT. Some vortices are still observed even when lowering the temperature without applying any field (0 T) because of the remaining residual fields in the surroundings. The successful demonstration of the SPM below liquid-helium temperatures and in non-zero magnetic field strengths ensures the nanoscopy platform our next step of cm-SNOM operation incorporating a THz light pulse to the tip.

We record approach curves with and without magnetic field by fixing at the time-domain delay that produces the highest near-field amplitude on a metallic surface. Fig.~\ref{fig:5} presents the data in a log-log plot at the sample temperature of 11.5 K and field strengths at 0 T and 5 T. As shown in Fig.~\ref{fig:5}(a), THz $s_{1}$ signals at zero field (black circles) decline relatively slowly with the tip-sample distance. We find that $s_{1}$ signal drops $\sim$20$\%$ from its maximum at a tip-sample distance of $\sim$5$\mu m$. Although the drop is still much less than the THz wavelength of 300$\mu$m, the relatively slow $s_{1}$ change can be understood from our appratus design as shown in Fig.~\ref{fig:2}(c).
To compromise with the compact space at the center of the magnet, we use a smaller parabolic mirror that 
causes a reduction in the numerical aperture for collecting near-field responses and an increase in the proportion of the far-field contribution in the scattered signals. 
In comparison, THz $s_{2}$ signals at zero field (red solid circles, Fig.~\ref{fig:5}(a)) drops sharply and we find the $\sim$20$\%$ decrease from the maximum at less than 200 nm tip-sample distance, indicative of a strong near-field component. It is interesting to note that the $s_{1}$ and $s_{2}$ signals exhibit similar time traces, as shown in Fig.~\ref{fig:3}(a) and (c), which indicates the near-field contribution still dominates the far field ones even in THz $s_{1}$ signals. 
Most intriguingly, applying magnetic fields allows a much sharper approach curve drop within $\sim$1$\mu$m even for THz $s_{1}$ signals which enhances near-field contributions as discussed below. 

Near-field signals at 5 T are smaller and approximately a third of that measured at 0 T in the data described in Fig.~\ref{fig:5}(b). We attribute this observation to eddy currents that flow through the mechanically resonating AFM tip in the presence of a magnetic field. This restrains the tip tapping movement such that the tapping amplitude and the demodulated near-field signal is diminished. Recordings of the tip resonance curve, which plots the tip tapping amplitude as a function of the tapping frequency, corroborates this idea as the curves are severely dampened under magnetic fields and, depending on the tip and how it is mounted to the AFM head, the resonance amplitude at 5 T can be two to more than ten times suppressed than the zero-field curve. A higher AC voltage can be applied to the dither piezo to excite the cantilever and raise the tapping amplitude back to the value that was measured without an applied field; this may not be viable if the damping becomes too harsh. Interestingly, we noticed the SNOM approach curves under magnetic field (Fig.~\ref{fig:5}(b)) are much steeper than its zero-field counterpart (Fig.~\ref{fig:5}(a)). For quantitative comparison, the decay distance (1/$e$) of the $s_{1}$ scattered amplitude from the maximum value is $\sim$560 nm at 0 T but is significantly reduced to $\sim$230 nm at 5 T. Although both can be considered as THz near-field signals, the magneto-THz $s_{1}$ signal surprisingly consists of more than $\sim$75\% of the total signal within the 500-nm distance, i.e., dominantly a near-field contribution. This is similarly shown for $s_{2}$. Although the noisy data here raises an ambiguity in the analysis of the THz $s_{2}$ signals, the effective decay length of $\sim$130 nm at 0 T shrinks down to $\sim$20 nm at 5 T. 
Consequently, the vertically confined evanescent near field has a different profile when the magnet is turned on, and the probing depth will be shallower with less contributions from the unwanted far-field scatterings at higher fields. This feature would be worth examining in future studies. Importantly, as the decay distance of 230 nm is already a thousand times shorter than the free-space wavelength at 1 THz, this shows that one may use the higher $s_{1}$ signals for analyzing the near-field measurements accomplished at 5 T rather than $s_{2}$ which improves signal-to-noise ratio. 

\begin{figure}[]
	\includegraphics[width=\linewidth]{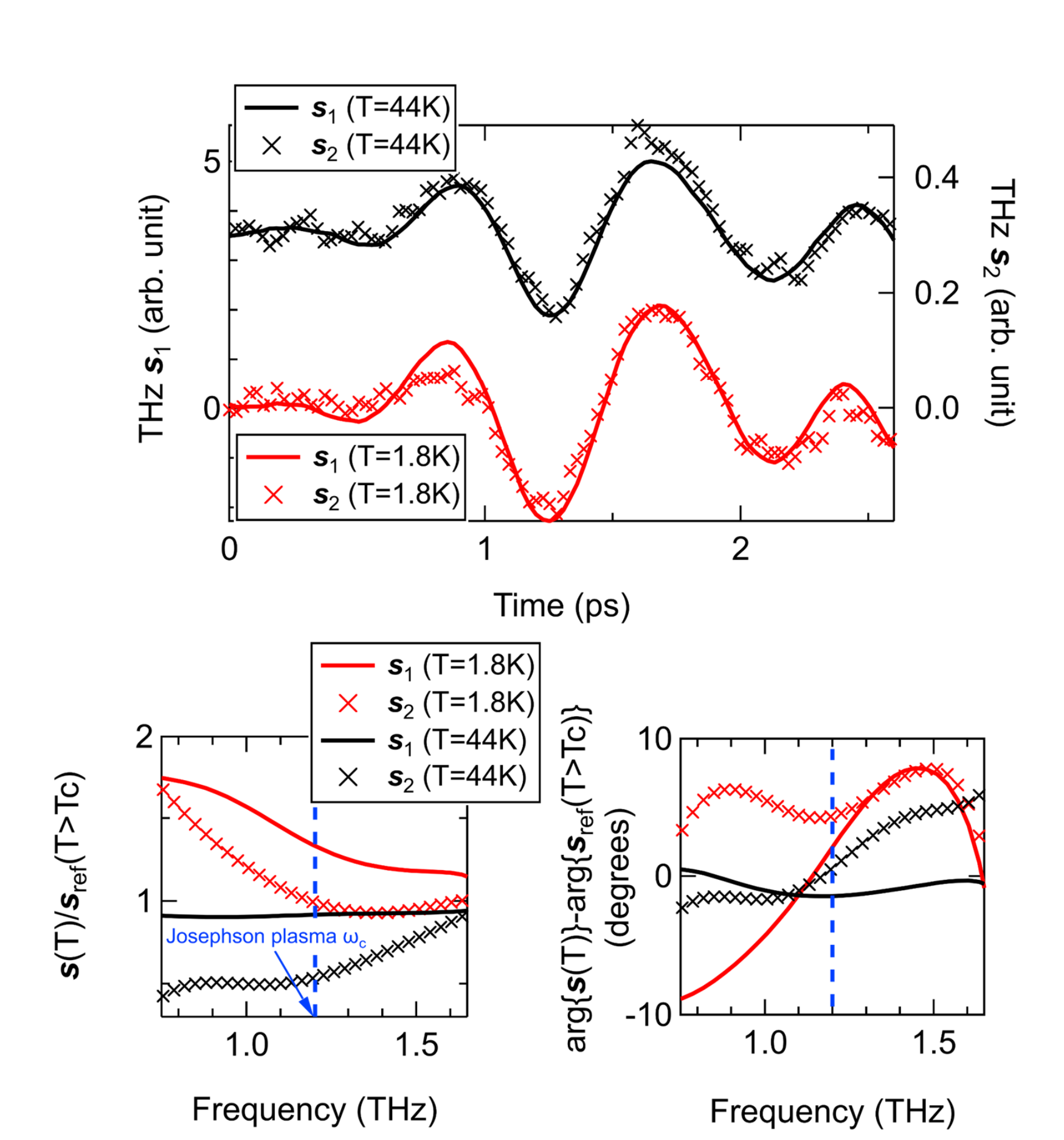}
	\caption{\label{fig:6} Cryogenic THzSNOM meausrement demonstrated on La$_{2-x}$Sr$_x$CuO$_4$ ($x$ = 0.17) at 1.8 K to detect Josephson plasma excitations. 
(a) Temporal waveforms of the THz tip scattering $s(t)$ at temperature 1.8 K (red line or cross) and 44 K (black line or cross). The THz-SNOM signals $s_{1}$ (solid line) and $s_{2}$ (cross) are plotted together. (b) Their Fourier-transformed spectral amplitude normalized with respect to the spectrum obtained from the time trace above $T_{\mathrm{c}}$. (c) Their Fourier-transformed spectral phase subtracted from the spectrum obtained above $T_{\mathrm{c}}$.}
\end{figure}

For operation of the THz cm-SNOM system below liquid-helium temperatures, we demonstrate a near-field detection of the distinguishing Josephson plasma excitations in a cuprate superconductor La$_{2-x}$Sr$_x$CuO$_4$ ($x$ = 0.17) as presented in Figure~\ref{fig:6}.  
The superconducting response results in the characteristic $c$-axis Josephson plasma edge of $\omega_{\mathrm{c}}\sim$1.2 THz at a temperature of $T$ $\ll$ $T_{\mathrm{c}}$$\sim$36 K.      
Having the single-crystal sample cut to expose the $ab$-plane, it should be emphasized that coupling between the THz light and the Josephson plasma becomes highly enhanced in our near-field setup geometry. 
This is due to the dominant near-field tip polarization along the sample normal, i.e., parallel to the $c$-axis Josephson plasma.
Fig.~\ref{fig:6}(a) presents the typical THz electric field waveform $\mathrm{s}(t)$ (inset) of tip scattering signals at 1$^{\mathrm{st}}$ ($s_{1}$, solid line) and 2$^{\mathrm{nd}}$ ($s_{2}$, cross) harmonics of the tip-tapping frequency. 
The Fourier-transformed spectral amplitude and phase obtained from the time-domain signals are shown in Fig.~\ref{fig:6}(b) and (c). The THz-SNOM spectra $s_{1}(\omega)$ at 1.8 K (red solid line), normalized by the reference spectra taken above $T_{\mathrm{c}}$, clearly show pronounced spectral plateau responses at $\omega_{\mathrm{c}}\sim$1.2 THz (blue dash line) while the $s_{1}(\omega)$ (black solid line) at 44 K does not show the Josephson plasma feature.  
The higher order $s_{2}(\omega)$ spectra exhibit the similar temperature dependent spectral behavior albeit having limited signal-to-noise ratio. 
These data present the first {\em near-field} spectra taken below liquid-helium temperatures.
Further experimental investigations to elaborate the physics will be published elsewhere as it is out of the scope of this initial instrument demonstration presented here.   

\begin{figure}[]
\includegraphics[width=\linewidth]{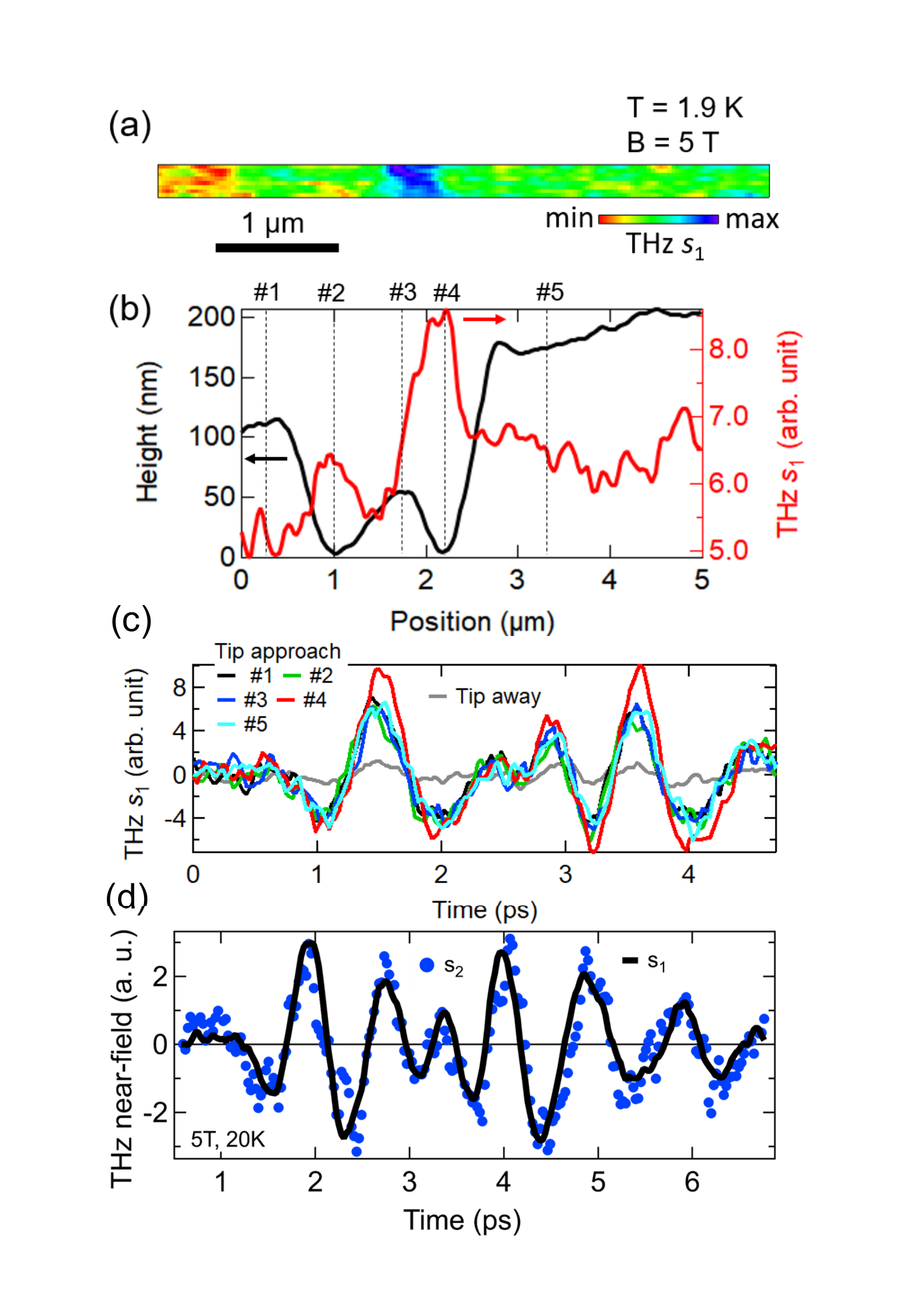}
\caption{\label{fig:7} THz cryogenic magneto-SNOM demonstrated on ZrTe$_{5}$ at 1.9 K and 5 T. (a) Spatial map of THz $s_{1}$. (b) Line profiles of the THz signal and sample height taken along the horizontal axis of (a). Positions $\#$1 through $\#$5 indicate where THz time traces are recorded. (c) Time-domain THz waveform of near-field scattering amplitude $s_{1}$ taken at five different positions.  Also shown together is time-domain THz waveform when the sample is retracted from the tip $\sim$ 5 $\mu m$ (gray trace in (c)). (d) Temporal profiles of tip-scattering THz signals are shown at 1$^{\mathrm{st}}$ ($s_{1}$, black line) and 2$^{\mathrm{nd}}$ ($s_{2}$, blue circle) harmonics of the tip-tapping frequency. 
}
\end{figure}

For operation of the THz cm-SNOM system under magnetic fields, we examined a topological semimetal, ZrTe$_{5}$, and demonstrated THz near-field measurements at a temperature and magnetic field of 1.9~K and 5~T. Fig.~\ref{fig:7}(a) presents a THz scan with the sample $c$ axis dominantly aligned along the horizontal axis of the image. Remarkably, the THz $s_{1}$ image clearly distinguishes sub-micron transition features from various domains and proves that the detected scattered signal indeed emanates from the near-field contribution from the tip apex and the sample underneath. Line cuts of the sample topography and the simultaneously taken THz near field along the horizontal axis are shown in Fig.~\ref{fig:7}(b). An increase in the scattering amplitude appearing at locations where there is a topographical change is normally expected. The near-field signal grows as the AFM tip approaches points $\#$2 and $\#$4, where it becomes surrounded by the surface below and the sides which increases the total area of contact with the sample. However, the two almost topographically identical trenches show a large deviation in the THz $s_{1}$ amplitude, which implies an electronic origin for the trough with the brighter near-field contrast. We further detected THz times traces on the different regions marked in Fig.~\ref{fig:7}(b). We chose five positions to obtain the electric-field waveform of the THz scattering response as a function of time as shown in Fig.~\ref{fig:7}(c). 
These measurements demonstrate cm-SNOM's capabilities for coherent THz detection under high magnetic field. Moreover, Fig.~\ref{fig:7}(d) compares the THz electric field waveform $\mathrm{s}(t)$ of tip-scattered signals at 1$^{\mathrm{st}}$ ($s_{1}$, black line) and 2$^{\mathrm{nd}}$ ($s_{2}$, blue circle) harmonics of the tip-tapping frequency. These two electric-field traces, taken at 5 T and 20 K, show very similar temporal profiles, with the exception that the higher order harmonics have smaller signal sizes. Since the $s_{2}$ THz signals exhibit a sharp drop in their approach curves (red solid circles in Fig.~\ref{fig:5}(a) and (b)), the waveform similarity between $s_{1}$ and $s_{2}$ indicates one may use the higher $s_{1}$ signals for near-field spectroscopy measurements, especially at high magnetic fields (black circles in Fig.~\ref{fig:5}(b)). Therefore, we conclude that both $s_{1}$ and $s_{2}$ signals dominantly measure the local magneto-THz responses in the vicinity of the tip-sample near-field region. 
Details on the magnetic-field-induced local coherent transport responses is beyond the scope of the instrument demonstration which  
will be published elsewhere.

In summary, we developed a one-of-a-kind cryogenic magneto-THz nanoscope, cm-SNOM, that offers both strong magnetic fields and below liquid helium temperature operation. We have conducted proof of principle near-field THz imaging and spectroscopy on cuprate superconductors and topological semimetals. 
Our development of cm-SNOM opens fascinating opportunities to access spatial inhomogeneity of nonequilibrium superconductivity by probing the THz gap energy, exotic low-energy bound states in superconducting vortex cores, or provide direct visualization of dissipationless edge conductions and their dynamics involved in phase transitions in diverse topological systems.

\begin{acknowledgments}
	This work was supported by the US Department of Energy, Office of Science,
Basic Energy Sciences, Materials Science and Engineering Division under contract No. DEAC02-
07CH11358 (THz study of topological materials). 
cm-THzSNOM Instrument was developed by the W.M. Keck Foundation (initial design and commission) and by the U.S. Department of Energy, Office of Science, National Quantum Information Science Research Centers, Superconducting Quantum Materials and Systems Center (SQMS) under the contract No. DE-AC02-07CH11359 (upgrade for improved cryogenic and magnetic operation).
We are grateful of various test samples from Ames National Laboratory (Y. Mudryk), Brookhaven National Laboratory (G. Gu and Q. Li) and Rigetti Computing for commission of the cm-THzSNOM.  
We are also grateful of various discussions from P. Canfield, Z. Fei, T. Koschny and C. Soukoulis at the initial stage of the project. 

\end{acknowledgments}

\section*{Data Availability Statement}

The data that support the plots within this paper and other findings of this study
are available from the corresponding author upon reasonable request.

\nocite{*}

\end{document}